# Reply to "Comment on 'Precision measurement of the Casimir-Lifshitz force in a fluid'"

J. N. Munday[1] and Federico Capasso[2]

[1]*Department of Physics, Harvard University, Cambridge, MA 02138*
[2]*School of Engineering and Applied Sciences, Harvard University, Cambridge, MA 02138*

We have reviewed the Comment of Geyer *et al.* [Phys. Rev. A, Preceding Comment] concerning our recent work [Phys. Rev. A **75**, 060102 (R) (2007)], and while we disagree with their criticisms, we acknowledge them for giving us the opportunity to add interesting addition material and a more detailed description of our experiment. We describe further our calculation and explain why a more sophisticated model is not warranted. We also present detailed experiments on the effects of electrostatic forces in our measurements and show that the contribution due to work function differences is small and that the residual electrostatic force is dominated by trapped charges and external fields. Finally, we estimate the effect of double layer interactions. These additional calculations and measurements support our original conclusion that the experimental results are "consistent with Lifshitz's theory."

**First**, we address the remarks [1] concerning the model/data for the dielectric functions used for the calculation of the Casimir force in our paper[2]. The model we used for gold was a Drude model $\varepsilon(\omega) = 1 - \frac{\omega_p^2}{\omega(\omega + i\gamma)}$, where $\omega_p = 9.0\,\text{eV}$ and $\gamma = 0.035\,\text{eV}$. We did not use the full frequency dependence of $\varepsilon(\omega)$, in contrast to the calculations of [1]. This is the reason of the non-agreement. We only used the Drude model because, in absence of a real measurement of the dielectric properties of the materials used in the experiment, a more sophisticated model for $\varepsilon(\omega)$ can not provide a "true" theoretical curve. We add here detailed calculations of the Casimir force using different models for epsilon and conclude that it can vary by 25% or more (depending on distance) and that our original calculation using the simple Drude model falls within this range of possible force curves.

In particular, using the same procedure as Ref[1], we have performed calculations that take into account the relatively large spread of published data for the dielectric function of gold (see Fig. 1). It is indeed widely known that the values of the dielectric properties of metallic films can vary in a significant range (~10-20%) due to different film preparation conditions. We (and all other groups performing Casimir force measurements thus far) have not reported measurements of the dielectric function for the *actual* films used in the experiments, which would have to be performed over an extremely wide frequency range for the gold thin films deposited on the plate and the sphere to test the accuracy and relative merits of different calculations for comparison to experimental force data as already pointed out by a number of authors [3-5].

Figure 1 shows our calculation of the Casimir force using the *same* methodology as Geyer *et al.* for obtaining the dielectric function from different Drude model parameters [3] and different optical data, taken either from Palik [6] (*dark red band*) or Weaver [7][8] (*light blue band*); the *light red band* is where the dark red and light blue bands overlap. The lower boundary of the force calculations in Fig. 1 corresponds to those of Geyer *et al.* It is therefore obvious that had they used different values for the dielectric function (dependent on the properties of different films reported in the literature) they would have obtained a significantly smaller value of the force. We emphasize that the data of Palik do not correspond to any single gold film but rather were assembled from different authors who performed measurements on different samples using different deposition processes in order to cover the frequency region quoted in Palik. Our original calculations (dashed curve) fall within the range of possible force curves obtained with a more refined model for the dielectric function. In light of the above analysis, there is therefore no reason to prefer one of these calculations over any other using the Lifshitz theory.

**Second**, the authors of the Comment claim that we have underestimated the attractive electrostatic force by a factor of 590; however, this is based on the assumption that electrostatic force is due only to chemical potential differences. Additional experiments presented in this Reply contradict this notion and indicates that the resulting electrostatic force in our experiments depends primarily on effects other than chemical potential differences. These include spatial variations of the work function due to grain boundaries in polycrystalline metal films, electrostatic fields caused by charges trapped at the grain boundaries, adsorbed species, [9-12] and external stray fields. In the following, we will describe how we (a) determined the origins of the electrostatic forces, (b) determined that the contribution of the electrostatic force to the total measured force is small and (c) estimated the residual electrostatic force.

We first consider the origin of the electrostatic forces encountered in real world experiments. The notion that residual potential difference, usually denoted as $V_0$, is nonzero between two films of the same metal (in contact) is itself a manifestation of the significance of surface and interface effects. The non-ideal interface between the gold and air/ethanol results in a $V_0$ which is not simply equal to the work function difference between the two metal surfaces in vacuum. This fact was pointed out in the original measurements of the Casimir force by Sparnaay [9], who noted that the difference in surface potentials "depends on the surrounding atmosphere and shows drift." Overbeek later noted in the first reliable and systematic measurements of the Casimir effect using a plate-sphere configuration in air [10] that "in spite of careful compensation for the Volta effect, other electrostatic effects influenced the measurement. Although the origin of these forces is not properly understood it has been possible to eliminate these disturbing effects too." Such effects are caused by non-uniform work functions, known as patch potentials, and trapped surface charge, which lead to residual potentials differences that depend on position and time [11, 12]. Finally it is worth noting that a decade of Casimir force measurements by a number of groups [13-17] (and references therein) have shown that $V_0$ has a large variability (typically in the tens of mV to ~0.5V range), in contradiction with the claims in Ref[1] that the variation is small. While it is often difficult to quantify the origin of the various contributions to the electrostatic forces, they in general result from work function differences, trapped charges and stray external fields. These forces behave differently when the environment is changed from air to ethanol. In the simplest case, electrostatic forces due to work function differences will be increased by a factor of $\varepsilon_{ethanol}$, the dielectric constant of ethanol, while the force arising from trapped charge and external fields will decrease by a factor of $\varepsilon_{ethanol}$ due to the induced dielectric polarization. Performing the additional experiments described below using the same configuration as the original report [2], we have confirmed that the electrostatic force due to work function differences is small and have estimated the residual electrostatic force to arise mainly from trapped charge and external fields. This is the reason we stated that the electrostatic force is reduced by a factor of $\varepsilon_{ethanol}$ [2], rather than increased by $\varepsilon_{ethanol}$ as is suggested by the authors of the Comment.

To determine the residual electrostatic forces in ethanol, the effect of grounding the sphere and/or the plate was first investigated. Figure 2 shows data using a cantilever and the measurement scheme described in our original paper [2], where the detector signal is proportional to the deflection of the cantilever and hence to the force by Hooke's law. Three different grounding situations were considered: (1) grounding both the sphere and plate, (2) grounding only the sphere, and (3) floating both the sphere and plate (no grounding). Note that because the force calibration technique relies on the hydrodynamic force rather than the electrostatic force [2], measurements of the cantilever displacement, and hence force, can be made with or without grounding the sphere and/or plate. The data (Fig. 2(a)) clearly show that for the three cases, the error bars (indicating sample standard deviations) are overlapping. To quantify the relevance of the observed deviations, we tested their statistical significance by applying an unpaired t-test [18] (according to Satterthwaite's approximation) with a two-sided alternate at two different separations (30nm and 55nm) for cases (1) and (3), which have the largest difference in means. We obtained P-values of 0.0788 and 0.4977 for separations of 30nm and 55nm, respectively; thus at 30nm, there is an $\approx 8\%$ chance of being wrong by rejecting the null hypothesis, while there is a $\approx 50\%$ chance of being wrong by rejecting the null hypothesis at 55nm. We therefore conclude that for these data, the difference in the means is not statistically significant at the 0.05 level; however, the difference is significant at the 0.1 level for the smallest separation, 30nm. Thus, we are not willing to reject the null hypothesis in our experiment with a high level of confidence. If we were to reject the null hypothesis and consider the possible difference in the means for the 3 cases to be significant, the difference is much smaller than calculated by Geyer *et al.* under the assumption that the electrostatic force measured in air arose only from charge transfer between the Au films due to work function differences (Volta effect) [1]. We thus conclude that the total deflection arises from the sum of the Casimir force and, possibly, small electrostatic forces which arise from sources other than work function differences. We expect therefore that when comparing the cantilever deflection in air to that in ethanol that the latter should be greatly reduced due to dielectric screening. The second set of experiments (Fig. 2 (b)), repeated many times with several spheres/plates, confirm this expectation. In all cases, the electrostatic interaction between the grounded plate and the sphere was greatly reduced upon insertion of ethanol (note: experiments performed without grounding the sphere and plate showed a similar reduction). In air the electrostatic force is large compared to the Casimir force for distances > 100nm; therefore the data of Fig. 2 (b) clearly show that the electrostatic force in the presence of ethanol is reduced, to a level comparable to the noise.

Because we have shown that a work function difference gives at most a small contribution to the force in ethanol at short range where the Casimir force is dominant (Fig. 2 (a)), we want to determine a worst case scenario for the contribution due to trapped surface charges and external fields at these distances. If we assume that the force resulting from the potential difference measured in air is due *entirely* to such charges and fields, the electrostatic force in ethanol would be reduced to $F_{ethanol} = F_{air}/\varepsilon_{ethanol}$. We note that the equation used on page 3 of Ref. [2] to estimate the residual force is not strictly correct because $F_{air} \neq -\pi R \varepsilon_0 V_0^2 / d$ due to the complexity of the resulting electrostatic force described above; however, the overall effect of the fluid is to reduce $F_{air}$ by $1/\varepsilon_{ethanol}$.

**Third**, we address the criticism in Ref[2] that we did not discuss the effect of *double layer forces* in Ref[2]. These were not mentioned because their effect is generally small unless additional salt ions are added to the relatively pure initial solutions. Furthermore, the effect of an ionic solution, as opposed to a pure solvent, is to further reduce the electrostatic contribution to the total force. The authors of the Comment cite Ref. [19] for experimental evidence of such double layers; however, the authors of Ref. [19] only report double layer forces in aqueous solutions and not in ethanol (note: additional experiments on double layer forces are discussed in Ref. [20] and references therein). Between conducting surfaces in ethanol, they conclude that "the small attractive force just before the tip jumps into contact *is due to van der Waals interaction.*" This does not mean that double layer forces are nonexistent in ethanol, but rather that the van der Waals force is large by comparison. It is well known that the van der Waals force is the non-retarded limit of the Lifshitz force for small separations, and thus, our results [2] are in agreement with the conclusions of Ref. [19].

To consider the possibility of double layer screening in our original experiment (note that no salt ions were added to the solution), we have calculated the approximate Debye screening length due to possible impurities by assuming that the residue left upon evaporation of the fluid (0.00036% by mass) [21] is from a common salt, NaCl as suggested by the authors of the Comment. This leads to a maximum concentration of a 48.6µM and a Debye length of $\lambda=24$nm [22]. This causes an additional screening of the electrostatic force in ethanol; however, it is not large enough to screen the force due to a $V_0$=130mV resulting from a work function difference, as suggested by Geyer *et al.*[1] The electrostatic force in ethanol would be $F_{ethanol} = -(\pi R \varepsilon_{ethanol} \varepsilon_0 V_0^2 / d) e^{-d/\lambda}$, the first factor accounting for the enhancement of the field upon immersion in the fluid, and would have a value of 1.2nN at d=40nm. This is approximately 5 times larger than the forces we measure in ethanol. Thus, this does not support the hypothesis of Geyer *et al.* that large electrostatic forces exist which are screened by the double layer. However, the fact that salt contaminants exist even in the purest available solutions means that *all* electrostatic interactions in our experiment are further screened.

In summary, from detailed additional measurements presented in this Reply, we conclude that electrostatic forces give negligible contributions in the range of distances of our measurements. Furthermore, Debye screening additionally reduces the electrostatic force. It should also be noted that the hydrodynamic force may play a greater role in the total residual force if the electrostatic force is *completely* screened. In this case, the residual force would be 12pN at 40nm. This corresponds to less than 5% of the total force at this separation, which is within the error bars shown in Fig. 3 of Ref. [2].

In conclusion, we have answered the questions raised by the authors of the Comment and have shown additional data to help clarify the origins of the residual electrostatic forces, which supports our original conclusions. The significance of our original paper [2] stems from it being the first report of the measurement of the Casimir force in a fluid and from verifying that it generally follows the Lifshitz theory. The latter is a non-trivial point due to the presence of dissipation in the dielectric, as recently pointed out to us [23], because "there is no general equation for the stress tensor in a dielectric with dissipation. Expression of the tensor in the terms of fluctuations is a non-trivial problem, which can be solved in the thermodynamic equilibrium only. After, of course, one uses the fluctuation dissipation theorem to calculate the fluctuations."

We are grateful to L. Pitaevskii for useful correspondence, to M. Romanowsky, A. Parsegian and G. Carugno for discussions, V. B. Svetovoy for providing optical data, and P. Wolfe for assistance in the statistical analysis.

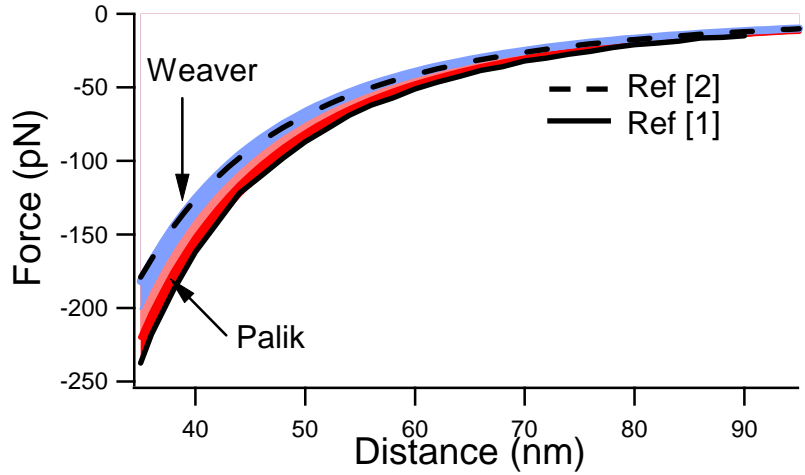

Fig. 1. (Color Online) Calculations of the Casimir force between a gold sphere and a gold plate immersed in ethanol using different data for the dielectric function, performed using the same method of Ref.[1] Solid bands correspond to possible values of the Casimir force resulting from different Drude parameters (determined from experiments on different samples) from Ref.[3] and from the optical data of Weaver (light blue) or Palik (dark red). The light red band corresponds to an overlap between both sets (Weaver and Palik). The calculation of Geyer et al. corresponds to the maximum magnitude of the force using Palik's data, while our original calculation from Ref.[2] is closer to that of Weaver's data.

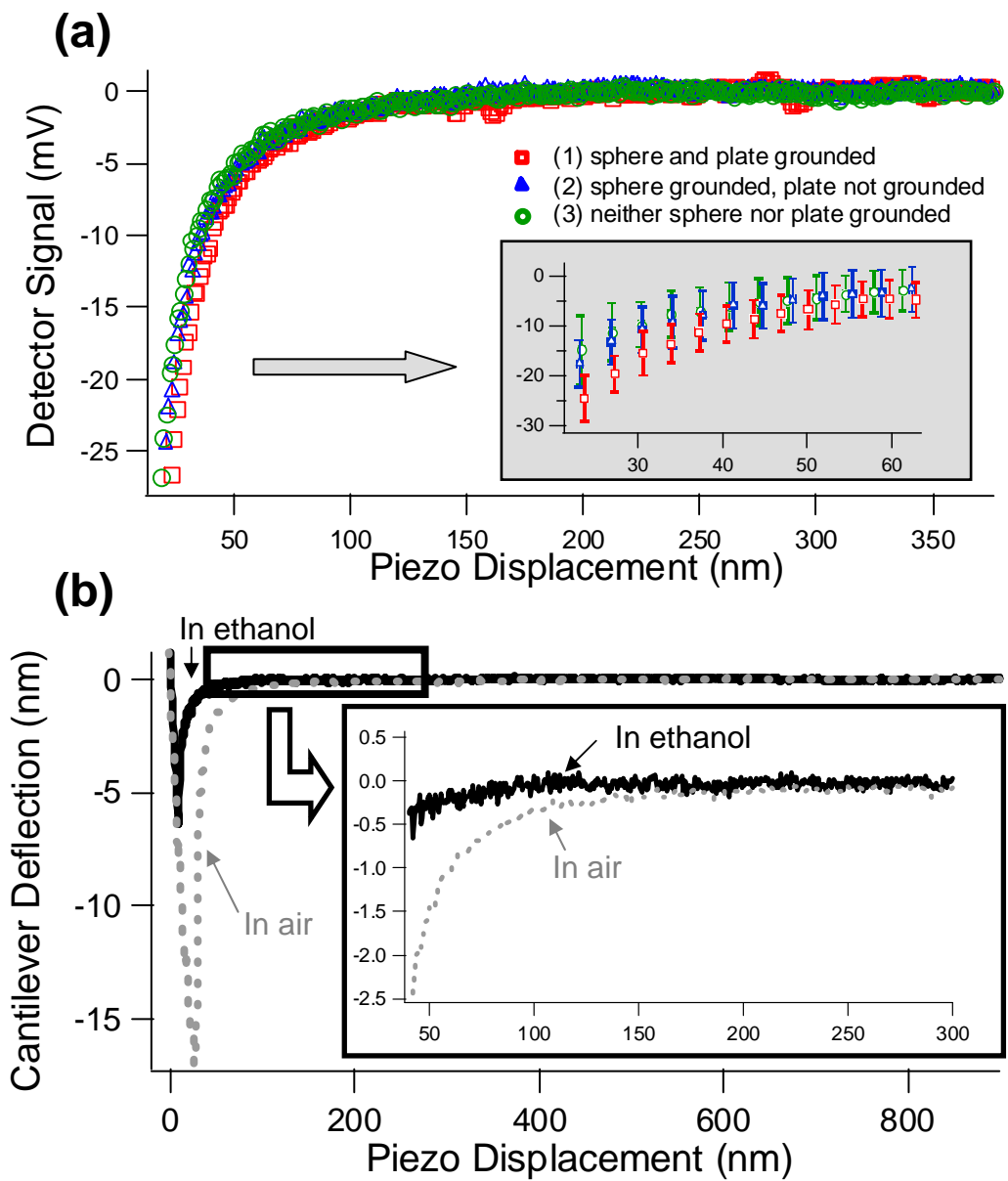

Fig. 2. (Color Online) Raw cantilever deflection data. The detector signal is proportional to the cantilever deflection and hence to the force between the Au metallized plate and sphere. Zero piezo displacement corresponds to contact between the sphere and plate with the cantilever undeflected. (a) Three different grounding options are considered: grounding both the sphere and plate (red squares), grounding just the sphere (blue triangle), and no grounding (green circle). Inset: Data for small separations showing the variation of the force between these different cases. Error bars represent the standard deviations from 5 measurements. Only 1/3 of the data points are shown for clarity. (b) Deflection in air (dashed grey line) and in ethanol (solid black line) for both the sphere and plate grounded. Inset: Cantilever deflection versus piezo displacement in the range of 40-300nm. The electrostatic force is dominant in air above 100 nm. The data shows that it is strongly reduced in ethanol compared to air. The data were reproduced with several spheres.